\documentclass{article}

\usepackage{jcappub}

\usepackage{amsmath, amssymb}
\usepackage{epsfig,graphicx,color,subfigure} 
\usepackage{indentfirst}
\usepackage{textcomp}
\usepackage{braket} 
\usepackage[]{caption} 
\usepackage{geometry} 
\usepackage{xcolor} 
\usepackage{tikz}

\graphicspath{	
	{./pics/}
}

\newcommand{\be}{\begingroup \begin{eqnarray} \begin{aligned}}
\newcommand{\ee}{\end{aligned} \end{eqnarray} \endgroup}

\newcommand*\xbar[1]{%
	\hbox{%
		\vbox{%
			\hrule height 0.4pt 
			\kern0.3ex
			\hbox{%
				\kern-0em
				\ensuremath{#1}%
				\kern-0em
			}%
		}%
	}%
} 
\newcommand{\mc}{\mathcal}
\newcommand{\vk}{\varkappa}
\newcommand{\mpl}{m_{Pl}}
\newcommand{\abnc}{\epsilon_{PBH}}
\newcommand{\cpl}{g}
\newcommand{\lagr}{\mc{L}}
\newcommand{\dist}{{\ell}}
\newcommand{\intvbl}{{\tilde q}}
\newcommand{\press}{\mathcal{P}}
\newcommand{\pd}{\partial}
\newcommand{\dod}{\tau_0}
\newcommand{\cdls}{{\tilde c}}
\newcommand{\hc}{\dagger}

\renewcommand{\exp}[1]{{\rm exp} \left\lbrace #1 \right\rbrace}

\title{Accretion in very early Universe and charge asymmetry}
\author[a]{Pozdnyakov N.\,A.}
\affiliation[a]{Department of Physics, Novosibirsk State University,
	Pirogova 2, Novosibirsk 630090, Russia}
\affiliation[a]{Institute for Nuclear Research of Russian Academy of Sciences, 117312 Moscow, Russia}
\emailAdd{pozdniko@gmail.com}
\date{\today}
\keywords{accretion, baryon asymmetry}
\toccontinuoustrue

\begin{document}

\abstract{Additional elementary species and primordial black holes are common candidates for dark matter. Their co-existence in the early Universe leads to accretion if particles are heavy. We solve equation of motion affected by expansion which enhances black hole growth rates. They depend upon particles freeze-out time rather than their mass. Taking into account friction we investigate recently suggested baryogenesis mechanism operating via scattering cross-section difference between particles and antiparticles on relativistic symmetric plasma. We find that asymmetry is accumulated at relatively small times which can be used to construct viable particle models.
}

\maketitle

\section{\label{sec-intro}Introduction}

Cosmological observables hint towards the Standard Model (SM) incompleteness. Galaxy rotational curves, cluster masses, and density fluctuations imprinted in cosmological microwave background (CMB) are usually explained incorporating the cold dark matter (CDM) paradigm with parameter density $\Omega_{CDM,0}h^2 = 0.12$~\cite{PDG}. Two popular options are elementary DM and primordial black holes though there are other possibilities such as gravity modification. The second hint is the baryon asymmetry of the Universe (BAU) magnitude~\cite{PDG},
\be
\eta_s \equiv \frac{n_{b} - n_{\bar b}}{s} = (8.75 \pm 0.23) \times 10^{-11}.
\ee
If asymmetry is to come from particle interactions, then: {\bf (i)} $B$-number non-conservation, {\bf (ii)} $C$ and $CP$ symmetries violation (CPV), {\bf (iii)} deviation from thermal equilibrium~\cite{Sakharov-1967} are required. Though, the SM provides $B$ and $L_\alpha$ numbers violation allowing to consider initial lepton number asymmetry transfer into baryon number by leptogenesis mechanism. It requires relevant temperatures to be $T> T_{EW} \sim 160$\, GeV. Further details can be found in~\cite{Bodeker-2020, Barrow-2022} and refs. therein. 

Primordial black holes (PBHs) can also participate in asymmetry production. The common recipe incorporates Hawking emission being either symmetric with asymmetry in particles decay~\cite{Dolgov-1980}, or asymmetric in presence of chemical potential originated from, e.g. non-minimal coupling to gravity~\cite{Hook-2014,Hamada-2016}. Another type of mechanism~\cite{Dolgov-2020} suggests particles and antiparticles inflow difference as a baryon asymmetry source. An illustrative example comes from $e^-$ and $p^+$ accretion~\cite{Bambi-2008} whose mobility difference appears due to significant mass ratio $m_p/m_e \sim 10^3$. Thus capture rates differ and primordial black hole's electric charge appears. Similarly, CPV induced difference in particles and antiparticles mobility causes baryon number to be either lost or hidden inside the PBH. It therefore provides baryon asymmetry even when $B$ is conserved at the level of elementary particles interaction. 

In such mechanism scattering cross-section and CPV effects are significant and depend upon specific model. In ref.~\cite{Ambrosone-2022} the authors suggest Yukawa interactions and evaluate cross sections difference in fermion on fermion scattering based on interference with a triangle one-loop diagram while in original proposal only estimation by order of magnitude was given. Though it provides specific working model such scattering process cross section is known to decrease with decreasing temperature. It is therefore interesting to consider other possibilities, for instance, within the same model fermion on scalar scattering where CPV comes from interference with bubble one-loop diagram. If $N_{2,3}$-particles masses are quasi-degenerate, $0<|m_2-m_3| \ll m_{2,3}$ then corresponding CPV magnitude can be resonantly enhanced that was studied for decay processes~\cite{Pilaftsis-2003} while link between decay and scattering was investigated in~\cite{Buchmuller-1997}.  

In what follows we consider both geometrical limit $\sigma = \cpl^4/m^2$ and fermion on fermion scattering. Since it incorporates Yukawa interaction we refer to introduced particles as to heavy neutral leptons preferring leptogenesis mechanism.  We choose $N_1$ particles to be the lightest, $m_1 < m_{2,3}$, and further drop index $1$. There is no straightforward correlation between resulting asymmetry magnitude and scattering cross sections since it also increases annihilation processes. Besides, larger cross sections also heavy particles prolong accretion proceeding in two-component liquid regime experiencing pressure from relativistic particles on motion. On the other hand, small cross-section also minimizes the mobility difference and is likely to provide dust energy density in conflict with observational data. Depending on chosen parameters one may find any combination of primordial black holes and particles, or none of them to represent dark matter. We do not further discuss available parameters space leaving it for the succeeding study.

We consider particles initially in thermal equilibrium. As usually, evolution operator $\hat L = \frac{\pd}{\pd t} - H p \frac{\pd}{\pd p}$ is simplified in $y \equiv p/T$ and $x \equiv m / T$ coordinates to $\hat L = Hx \frac{\pd}{\pd x}$. Moreover, dark matter evolution is usually described in terms of $Y \equiv n/s$ with entropy density,
\be
s \equiv \frac{\rho_{rel} + \press_{rel}}{T} = \frac{2\pi^2}{45} g_* T^3,
\ee
since $Y$ does not depend on time in isentropic Universe. $T=T_\gamma$ is the temperature of photons. We adopt this formalism and also normalize distances by black hole radius at the accretion beginning $r_g^* = r_g(t_{in})$. Therefore we introduce dimensionless $(x, \dist)$-coordinates
\be
t = \frac{x^2}{2m_*},~~~~~r = \dist \, r_g^*,
\label{3-dls-vars}
\ee
with expansion scale $m_* \equiv H x^2 = m^2 / \mpl^*$, where $\mpl^*(T) \equiv \sqrt{\frac{90}{8\pi^3 g_*(T)}} \mpl \sim 7 \times 10^{17}\,{\rm GeV}$ only weakly depending on temperature and additional to the SM particles. 
We note that since $x\propto\sqrt{t}$ velocity in such coordinates changes with time, $v = {\pd r}/{\pd t} = x_a^2\pd \dist / x\pd x$, therefore sound speed $\cdls_s$ and speed of light $\cdls$ also change in time.

Accretion in the early Universe is studied in the literature mainly in connection with PBHs mass growth (see, for e.g.~\cite{Custodio-2002, Ricotti-2007, Bisnovatyi-Kogan-2008, Mahapatra-2013}). However, accretion is shown to be able to locally generate electric charge asymmetry in present Universe~\cite{Bambi-2008,Dolgov-2023}. Accretion in the early Universe is influenced by stronger expansion rates modifying asymmetry evolution, however only an estimation is provided in~\cite{Dolgov-2021}.

In the following section we discuss primordial black holes. Then in section~\ref{sec-pcl} we discuss particles and their residual number density. Equation of motion and corresponding solution are presented in section~\ref{sec-mot-eq}. Then we describe accretion in the early Universe, concluding in section~\ref{sec-disc}.

\section{\label{sec-pbh}Primordial black holes}

Black hole appearance in the early Universe requires energy density perturbations above threshold $\delta_c \sim 1/3$. There are many relevant mechanisms providing different mass spectrum~(for details see ref.~\cite{Carr-2021}). An approximate monochromatic is assumed herein,
\be
\frac{dn}{dM} = \abnc \frac{2.7T_a}{M} n_{rel} \delta(M - M_0),
\label{2-PBH-spectrum}
\ee
with an initial abundance $\abnc \equiv \rho_{PBH}/\rho_{rel} \ll 1$ at appearance time $t_a$. It might be justified if PBHs appear due to equation of state softening assosiated with $N_{2,3}$-particles transition to non-relativistic. We also assume initial mass $M_0$ to be about of horizon,
\be
M_0 \simeq \mpl^2 t_a = \sqrt{\frac{90}{32\pi^3g_*(T_a)}} \frac{\mpl^3}{T_a^2} \approx 0.85 \times 10^7 \left(\frac{x_a}{3}\right)^2 \left(\frac{10^{13}\,\rm GeV}{m}\right)^2\,{\rm g},
\label{2-M-connection}
\ee
if $T_a > T_{EW}$, however pressure gradient effect~\cite{Nadezhin-1978} may decrease $M_0$. Characteristic distance between neighbouring primordial black holes is $\displaystyle \lambda = n_{PBH}^{-1/3} \approx 2 \abnc^{-1/3} x_a x/m_*$ given even spatial distribution. Then clustering processes start from $r_H \simeq \lambda$ with corresponding time $x_{cl} = \abnc^{-1/3} x_a$. Besides, initially small angular momentum~\cite{Harada-2021} can increase spoiling spherical symmetry and diminishing accretion rates.

PBH evolution before non-relativistic particles accretion is governed by evaporation and relativistic species absorption,
\be
\frac{\pd M}{\pd t} = -\alpha(M)\frac{\mpl^4}{M^2} + 27\pi \rho_{rel}\frac{M^2}{\mpl^4},
\label{2-early-M-evol}
\ee
with relativistic particles capture cross-section $\sigma_{rel} = 27\pi G^2M^2$ and $\alpha(M) \sim 5 \cdot 10^{-6} \, [7.8 D_{1}(M) + 3 D_{1/2}(M)] \lesssim 2.4 \cdot 10^{-3}$ with quantum number multiplicities $D_i = {\rm Spin} \times {\rm Charge} \times {\rm Color}$. Evolution is sharply dominated by one term: $\dot M_{abs} / \dot M_{evap} = (M/M_c)^4$, with
\be
M_c \approx 0.03 \frac{\mpl^2}{T} = \frac{0.9 \cdot 10^{-6}}{x} \, \frac{m}{10^{13}\,\rm GeV} M_{hor}
\label{2-crit-mass}
\ee
If absorption can be disregarded PBH is expected to evaporate at 
\be
x_{e} = x_a \sqrt{ 1 + 5 \cdot 10^{22} \left(\frac{x_a}{3}\right)^2 \left(\frac{10^{13}\,\rm GeV}{m}\right)^2}
\ee
In other limit eq.~(\ref{2-early-M-evol}) has solution,
\be
M = M_0 \left[1 - K \frac{M_0}{\mpl^2 t_a} \left( 1 - \frac{t_a}{t} \right)\right]^{-1},
\ee
with $\rho_{rel} = 3 \mpl^2/32\pi t^2$ $K = 81 / 32$ in agreement with ref.~\cite{Bisnovatyi-Kogan-2008}, however different $K \sim 0.04$ is found in ref.~\cite{Custodio-2002}, probably due to different $\rho_{rel}$ definition. Also, numerical study~\cite{Lora-Clavijo-2013} supports the possibility of significant mass growth due to radiation absorption. On the other hand, rotating black holes accretion rate~\cite{Mahapatra-2013} is smaller while evaporation time is larger. 

To avoid these uncertainties of PBH mass increase due to radiation accretion we consider $x_a / x_{in} > \sqrt{1 - \mpl^2 t_a / K M_0} \sim 0.78$ for the discussed $\rho_{rel}$. If $K < 1$ then mass gain is bounded by $(M-M_0)/M_0 < K$ and PBH enters evaporation dominated regime at
\be
x_* \sim \left( K \frac{M_0}{\mpl^2 t_a} + 1\right) \frac{M_0}{0.03\mpl^2} m  \lesssim 2 \cdot 10^7 \left(\frac{x_a}{3}\right)^2 \frac{10^{13}\,\rm GeV}{m}.
\ee
therefore condition on $x_a/x_{in}$ is softened.

Then $M(t_{in}) \simeq M(t_a) \sim M_{hor}(t_a)$ i.e. $r_g(t_a) \sim r_H(t_a)$ and cosmological horizon ($q > 0$)
\be
r_H = a(t) \int_{0}^t \frac{dt'}{a(t')} \sim \frac{1}{qH} = \frac{x^2}{m_*},
\label{2-part-hor}
\ee
or $\dist_H = \left(x / x_a\right)^2$. Last equality in~(\ref{2-part-hor}) holds for radiation-dominated expansion since $t=0$.

\section{\label{sec-pcl}Heavy particles in the early Universe}

Our main focus is asymmetry generation mechanism proceeding via different particles and antiparticles absorption rates by primordial black holes. It can be treated either as baryogenesis with asymmetry transfer via decay or as leptogenesis. In ref.~\cite{Ambrosone-2022} Yukawa interactions,
\be
\lagr_{int} = -\cpl_{\bar aX} \phi^\hc \bar a X - \cpl_{\bar cX} \phi^\hc \bar c X - \cpl_{\bar bY} \phi^\hc \bar b Y - \cpl_{\bar Y X} \psi \bar Y X - \cpl_{\bar b a} \psi \bar b a - \cpl_{\bar b c} \psi \bar b c + {\rm h.c.},
\label{4-Yukawa-lagr}
\ee
were studied in context of baryogenesis. We find more natural to treat introduced $X,Y$ fields as heavy neutral leptons and therefore consider leptogenesis.

Since asymmetry is proportional to number of captured $N_1$-particles we consider them initially in thermal equilibrium with
\be
n_{eq} = \int \frac{d^3p}{(2\pi)^3} f_{eq}(p) = \frac{m^3}{2\pi^2 x^3} \int_0^{\infty} \frac{y^2 dy}{1 + e^{\varepsilon+\xi}}
\label{3-n-eq}
\ee
for $\xi \equiv \mu / T \sim 0$ and $\varepsilon \equiv E / T = \sqrt{y^2 + x^2}$ with $y \equiv p/T$. Starting from $x \sim 1$ annihilation processes dominate and number density changes according to
\be
\frac{\pd n}{\pd t} + 3Hn = -\braket{\sigma v} (n \bar n - n_{eq} \bar n_{eq}).
\ee
Cosmological term is absent in $(x,y)$-coordinates,
\be
\frac{\pd Y}{\pd x} = - \frac{\dod}{Y_{rel} x^2} (Y^2 - Y_{eq}^2),
\label{3-Y-evol}
\ee
with introduced $Y_{rel} \equiv {45\zeta(3)}/{2\pi^4}$, equilibrium yield
\be
Y_{eq} = \frac{g}{g_*}	\frac{45}{2\pi^4} \begin{cases} \zeta(3),~~x \ll 3, \\ \sqrt{\frac{\pi}{8}} x^{3/2} e^{-x},~~ x \gg 3, \end{cases}
\label{3-eq-yield}
\ee
and
\be
\dod \simeq 2.3 \cdot 10^{23} \sqrt{\frac{g_*}{106.7}} \frac{\braket{\sigma v}}{10^{-36}\,{\rm cm}^2} \frac{m}{10^{13}\,{\rm GeV}} = 93 \sqrt{\frac{g_*}{106.7}} \left( \frac{\cpl}{0.1}\right)^4 \frac{10^{13}\,\rm GeV}{m},
\label{3-dod-def}
\ee
in last equality we use geometrical cross-section, $\braket{\sigma v} = \cpl^4/m^2$.  

\begin{figure}
	\centering\includegraphics[width=.6\textwidth]{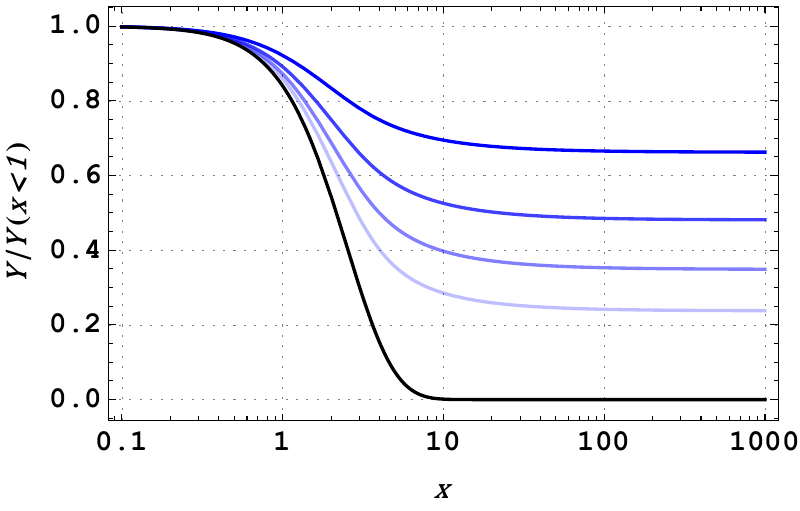}
	\caption{Evolution of equilibrium yield (black line) and numerical solution for $\dod = 100, 250, 500, 1000$.}
	\label{A-yields_1}
\end{figure}

Thermal dark matter features freeze-out mechanism because annihilation rates decrease faster than Hubble parameter in cooling Universe. After $x \sim \sqrt{\dod}$ (or $\braket{\sigma v}n \sim H$) $Y$ slowly declines from $Y_{eq}$, $Y - Y_{eq} \propto x^2 / \dod$. Once $Y \gg Y_{eq}$ an approximate solution of eq.~(\ref{3-Y-evol}) is
\be
Y(x) = Y_0 \left[ 1 + \frac{\dod}{x_0} \frac{ Y_0}{Y_{rel}} \left(1 - \frac{x_0}{x} \right)\right]^{-1}.
\label{3-Y-sol-fo}
\ee
Particles number density is frozen, $Y(x\to\infty) = Y_{fo} \sim Y_0$, if $x_0 \gg \dod Y_0 / Y_{rel}$ so characteristic time is $x_{fo} \sim \dod Y_{fo} / Y_{rel}$.

\paragraph{Gravity} In gravitational field determined by mass $M$ potential energy $U = -GMm/r$ is added to chemical potential changing equilibrium number density,
\be
n_{eq}'(r) = C n_{eq} \exp{\frac{GMm}{r T_1}}  = C n_{eq} \exp{\frac{x \dist_g}{2z\dist}}
\label{3-neq-BH}
\ee
where $C  = \exp{-x_a^2\dist_g / 2zx}$ is determined by $n_{eq}'(\dist_H) = n_{eq}$ and $T_1 \equiv z T_\gamma$ such that $z$ accounts for heating and cooling effects. Once such effect is caused by departure time from kinetic equilibrium. While particles energy is comparable to that of relativistic background collisions may lead either to energy increase and decrease. Therefore to maintain kinetic equilibrium $N \sim (T/\Delta E)^2 \sim x$ collisions are required with characteristic energy transfer $\Delta E \sim \sqrt{\omega^3/m}$. After $x_k \sim \sqrt{\dod}$ $N_1$'s temperature starts to evolve as $T_1~\propto~a^{-2}$ i.e. $z \equiv T_1/T = T/T_k = x_k/x$, for $x>x_k$.

Even after $x_{fo}$ particles number density can be changed in overdensity region near the black hole. Since afterward we adopt Lagrangian coordinates $Y = Y(x, \dist_0)$ we consider time derivative in eq.~(\ref{3-Y-evol}) as {\it individual} therefore
\be
\frac{\pd Y}{\pd x} = Y_\infty \frac{\pd \phi}{\pd x} - \frac{\dod}{Y_{rel}x^2} \left[Y^2  - e^{\frac{x^2 - x_a^2}{z x \dist}} Y_{eq}^2 \right].
\ee
where particles overdensity $\phi(x, \dist_0)$ is defined in~(\ref{4-ndens-profile}). We factorize $Y(x, \dist_0) = Y_\infty \phi(x, \dist_0) \Theta(x, \dist_0)$ therefore,
\be
\frac{\pd \Theta}{\pd x} = (1 - \Theta) \frac{\pd}{\pd x}\ln \phi - \frac{Y_\infty}{Y_{rel}} \frac{\dod\phi}{x^2} \Theta^2 + \frac{\exp{\frac{x^2 - x_a^2}{z x \dist}}}{\phi} \frac{\dod}{x^2} \frac{Y_{eq}^2}{Y_\infty Y_{rel}},
\label{3-ann-factor}
\ee
with initial condition $\Theta(x_0, \dist_0) = 1$. Considering quasi-steady flow at $x > x_k$, closest region $\dist < x^2 / \sqrt{\dod}$ is not significantly affected by annihilation due to exponential factor, but number density at larger distances can be decreased. Annihilation processes are important in a sense that they locally restore chemical equilibrium with the SM sector particles allowing to transfer accumulated asymmetry. After approximately $x \sim \phi(x_{cap}, \dist_0) x_{fo}$ asymmetry transfer ends.

\section{\label{sec-mot-eq}Equation of motion}

Emission redshift $z \sim H_0 r$ from distant galaxies is interpreted as uniform space stretching characterized by the Hubble parameter $H \equiv \dot a/a$. Today it is relatively small, $H_0 = 100\, h ~ {\rm km} \, {\rm Mpc^{-1} \, s^{-1}}$ with $h = 0.674 \pm 0.005$ from CMB, or $h = 0.73 \pm 0.01$ from SN\,Ia and Universe at the scales below $r = 200$ Mpc is inhomogeneous with gravitationally bound systems departing from {\it Hubble flow}. It is therefore argued whether expansion affect them at all scales or only at sufficiently large ones~\cite{McVittie-1933,Anderson-1995,Cooperstock-1998,Price-2005,Mashhoon-2007,Nandra-2011,Mars-2013,Jayswal-2024}. Though they agree on absence of observable effects below the scale of cluster of galaxies.

Fortunately, we consider early Universe consisting of homogeneous relativistic liquid with larger Hubble parameter. To get an idea of expansion influence consider $r = a(t) \hat r$ with co-moving radius, $\hat r$. Putting $\dot{\hat r} = 0$ recession velocity is $\displaystyle v_{exp} = \dot a \hat r = Hr = r / (1+q)t$ with deceleration parameter ($q \neq -1$),
\be
q \equiv -\frac{\ddot a a}{\dot a^2} = {- \frac{H'}{H^2} - 1} = \frac{(1+3w)}{2}.
\label{3-dec-par}
\ee
With coordinate change $\delta r \sim v_{exp} \delta t$ recession velocity becomes $v_{exp}'$ and
\be
F_{cosm} = \frac{v_{exp}' - v_{exp}}{\delta t} = \frac{-q}{1+q} \frac{v_{exp}}{t} = -qH^2r.
\label{3-cosm-force}
\ee
Determined this way it does not suggest causal connection between particles and only manifests the way Universe is expanding. In radiation-dominated stage $q=1$ and this force acts as an {\it inward} force. 

Motion of a test particle in presence of both gravity and expansion is described by McVittie metric~\cite{McVittie-1933}. We provide it in locally inertial frame corresponding to observer moving along with matter~\cite{Nandra-2011}
\be
ds^2 = \left[ 1 - \frac{2GM}{r} - r^2H^2\right] dt^2 + \frac{2rH}{\sqrt{1 - \frac{2GM}{r}}} drdt - \left( 1 - \frac{GM}{r}\right)^{-1} dr^2 - r^2 d\Omega^2.
\ee
In physical region $\dot r, r_g/r, rH \lesssim 1$ and it is useful to introduce $\vk \equiv r_g H = (x_a/x)^2 < r_g/r, rH \ll 1$. Then geodesic equation,
\be
\ddot r + \frac{r_g}{2r^2} + qH^2 r + \left( [1-q] rH - \frac{r_g}{r}\right) \frac{\vk}{2} - \frac{r_g/r - 2r^2H^2}{2} \left( \frac{\dot r}{\sqrt{1-r_g/r}} - Hr \right)^2 - \frac{q\vk^2}{8} = 0,
\label{3-geodesic-approx-eq}
\ee
for angular momentum $L=0$. Note that it does not contain cosmological friction term.

\begin{figure}
	\centering
	\includegraphics[width=.4\textwidth]{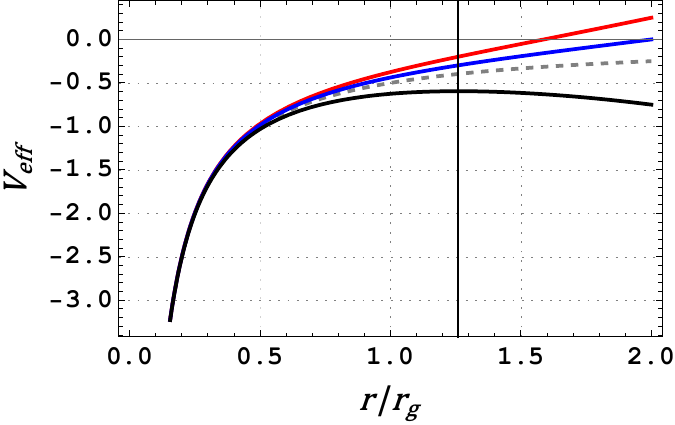}
	\includegraphics[width=.4\textwidth]{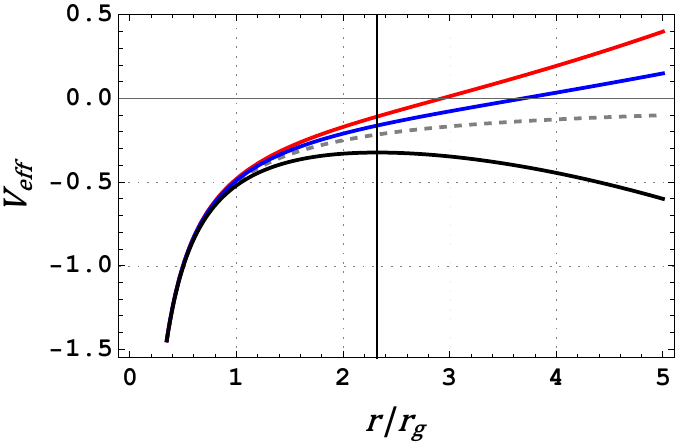}
	\includegraphics[width=.4\textwidth]{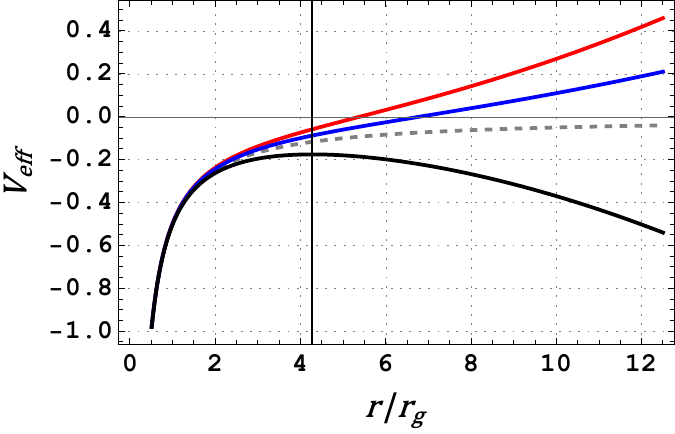}
	\includegraphics[width=.4\textwidth]{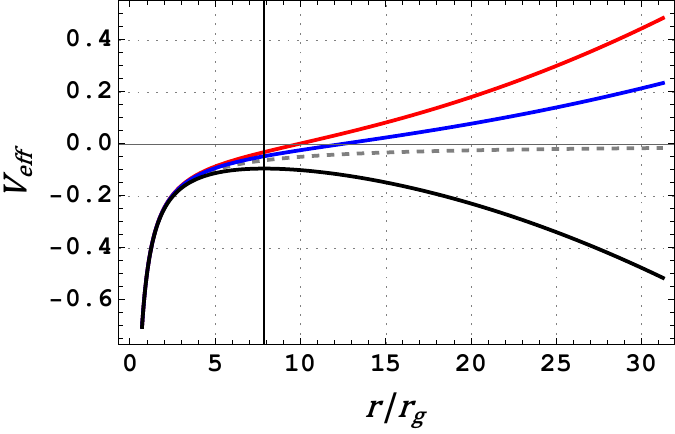}
	\caption{Effective potential $V_{eff}$ for deceleration parameters $q = 1$ (blue), $q=1/2$ (red), $q=-1$ (black). Dashed gray line represents only gravitational potential, while black vertical line marks $r_e(t)$. Pictures are provided for $\vk \equiv r_g/r_H = 0.05, 0.02, 0.008, 0.0032$.}
	\label{3-eff-pot-figure}
\end{figure}

The second and third terms in eq.~(\ref{3-geodesic-approx-eq}) equilibrate at distance
\be
r_e = r_e(t) = \left( \frac{|q|}{2} r_g r_H^2\right)^{1/3} = \left( \frac{q^2 \vk}{2}\right)^{1/3} r_H,
\label{3-eq-rad}
\ee
since~(\ref{2-part-hor}) and corresponding dimensionless distance, $\dist_e = (q^2/2)^{1/3} (x/x_a)^{4/3}$. It is drawn as vertical line in Figure~\ref{3-eff-pot-figure} for $V_{eff} = -{r_g}/{2r} + \frac{q}{2} H^2 r^2$ with different $\vk \sim r_g/r_H$ values. $V_{eff}$ can be view as potential evolution in time. Note, that the second term reminds kinetic energy associated with recession velocity.

Besides, we consider friction term $\gamma \dot r$ originated from scatterings on relativistic particles with $\gamma \equiv \braket{\sigma v} n_{rel} = 1 / r_{free}$ and the latter is 
\be
\dist_{free} = \frac{1}{\braket{\sigma v}} \frac{\pi^2}{g_* \zeta(3)} \frac{x^3}{m^3} \frac{m_*}{x_a^2} = \frac{x}{\dod(x)} \dist_H,
\label{3-mean-free-path}
\ee
where we have expressed thermally averaged scattering cross section $\braket{\sigma v}$ via introduced~(\ref{3-dod-def}). It is therefore also characterizes time $x_f = \dod$ when mean free path exceeds the cosmological horizon resulting in heavy particles freely propagating throughout the Universe. Time dependence $\dod = \dod(x)$ is assumed to be either constant with time, or proportional to $x^{-2}$.

Including Universe expansion and scattering processes, we consider equation of motion in form
\be
\ddot r + \gamma \dot r + q H^2 r + \frac{r_g}{2 r^2} = 0,
\label{3-motion-equation}
\ee
applicable in $r_g \ll r \ll r_H$ limit. We apply discussed motion equation of test particle to infinitesimally small volume that still contain many of $N_1$-particles, referred to as a liquid particle. In $(x,\dist)$-space with $q=1$ it takes form $x^2 \dist'' + \left( \dod - 2x \right) \dist' + \dist + \left({x}/{x_a}\right)^4 \, {\dist_g}/{2\dist^2} = 0$ and in limit $\dod \to 0$ it can be reduced to Emden-Fowler equation
\be
w'' = -(10\xi_a^{4\alpha})^{-1} \frac{\xi^{-\frac{1}{2}\left( 1 + \alpha \right)}}{w^2}
\ee
by means of $x = \xi^{\alpha}$ and $\dist = \xi^{(3\alpha+1)/2} w$ transformation, with $\alpha = 1/\sqrt{5}$. According to ref.~\cite{ZaiPol-2003} it has no exact solution.

To construct an approximate solution we incorporate $H \sim {\rm Const}$ approximation that requires $\left|{\delta H}/{H}\right| = {\delta t}/{t} \ll 1$ or equivalently $t_{cap} \ll 1/2H = t_H$. We normalize distances by $r_e(t_0)$~(\ref{3-eq-rad}) instead of $r_g$ with $0 \leqslant t \leqslant t_{cap} \ll t_0$ finding
\be
\ddot\rho + \gamma_0 \dot\rho + \frac{1}{4 t_0^2} \left( \rho + \frac{1}{\rho^2}\right) = 0.
\ee
We impose initial conditions $r_0 = r(t_0)$ and $v_0 = \dot r(t_0) = 0$ thus $\rho'(0) = 0$. Sometimes a different initial condition $v_0 = Hr_0$ is used~\cite{Mack-2006} which we believe not to be the case as is discussed later. Zeroth-order solution,
\be
t^{(0)} = -\int_{\rho_0}^\rho \frac{\sqrt{2}t_0d\rho}{\sqrt{\frac{1}{\rho} - \frac{1}{\rho_0} + \frac{\rho_0^2}{2} - \frac{\rho^2}{2} }} = \sqrt{2}t_0 \rho_0^{3/2} \int_{\rho/\rho_0}^{1} \frac{d\intvbl}{\sqrt{\frac{1}{\intvbl} - 1 + \frac{\rho_0^3}{2}(1 - \intvbl^2)}},
\ee
can be approximately integrated for $\rho_0 \lesssim 1$ as
\be
t^{(0)} & \approx \sqrt{2}t_0 \rho_0^{3/2} \int_q^1 \frac{ d\tilde q}{\sqrt{\frac{1}{\intvbl} - 1}} \left( 1 - \frac{\rho_0^3}{4} \left[ \intvbl + \frac{\intvbl^2}{4} - \frac{3\intvbl^3}{8} + \mc{O}(\intvbl^4)\right]\right) \\ & = \sqrt{\frac{r_0^3}{r_g}} \left[P_1 f_1\left(\frac{r}{r_0}\right) + P_2 f_2\left(\frac{r}{r_0}\right)\right],
\label{3-src-reg-sol}
\ee
with $\displaystyle P_1 \equiv 1 - 0.18 \frac{r_0^3}{2r_g t_0^2} - 0.12 \frac{r_0^2 r}{2r_g t_0^2} + 0.0065 \frac{r_0 r^2}{2r_g t_0^2} + 0.023 \frac{r^3}{2r_g t_0^2}$, $\displaystyle P_2 \equiv 1 - 0.18 \frac{r_0^3}{2r_g t_0^2}$ up to third order of Taylor expansion and introduced
\be
f_1(q) \equiv \sqrt{q-q^2}, \qquad f_2(q) \equiv {\rm arctg} \sqrt{1/q-1}.
\ee

First, in present Universe, i.e. $t_0 \to \infty$, or $r_H =\frac{1+q}{q} ct_0 \to \infty$ these coefficients are $P_i = 1$ and we come to well-known solution for radial motion in gravitational field. Second, if particle starts moving from $\rho_0 = 1$ or $r_0 = r_e(t_0)$ then $\displaystyle H t_{cap} \sim \frac{1}{\sqrt{2}} 0.82 \frac{\pi}{2} \sim 0.91$. Hence approximation $H = {\rm Const}$ breaks at $r_0 \gtrsim r_e(t_0)$ and this region should be treated differently. For instance, substitution $\rho = r/r_e(t)$ leads to
\be
t^2 \ddot \rho + \left( \frac{4}{3} + \gamma(t) t\right) t \dot \rho + \left( \frac{1}{36} + \frac{2}{3} \gamma(t) t\right)\rho + \frac{1}{4\rho^2}=0.
\ee
In limit $\gamma \to 0$, gravitational term can be disregarded at initial distances $r \gtrsim 2 r_e$ yielding solution $\rho(t) = t^n$ with $n = -1/6$ or $r(t)\propto t^{1/2}$. These particles join the Hubble flow. As pointed out in~\cite{Dolgov-2021} friction changes the move-away distance. In the intermediate region, $r_e < r < 2 r_e$ we assume particles motion to change from attraction to repulsion and therefore can be broadly considered to be in rest. Hence we find $v_0 = 0$ as a proper initial condition. Since $r_{e}(t)$ increases with time more particles appear in close region and begin motion towards BH. 

\begin{figure}
	\centering\includegraphics[width=.45\textwidth]{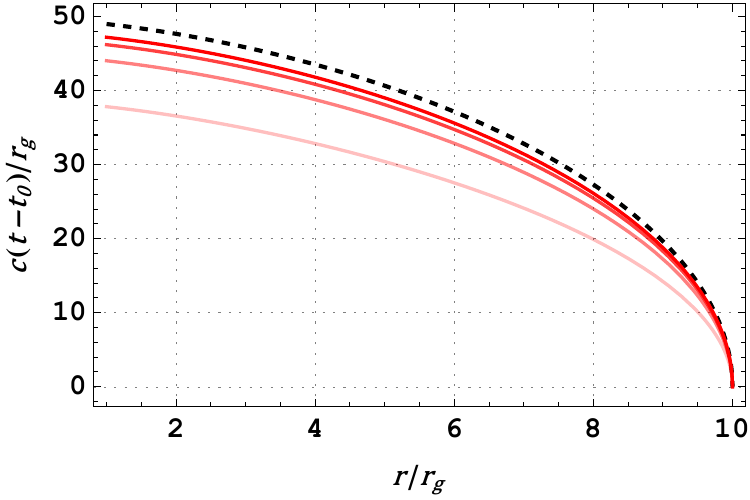}\includegraphics[width=.45\textwidth]{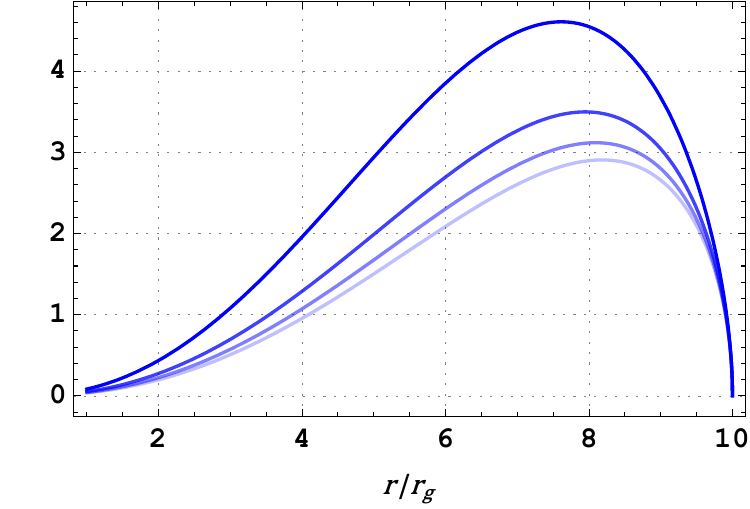}	
	\caption{Left: black line represents free fall in gravity, red lines correspond to~(\ref{3-src-reg-sol}) for $ct_0/r_g = 20,30,40,50$. Right: second term~(\ref{3-fric-sol}) integrand $e^F {w^{(0)}}^2$ for $x_a = 1$, $\dod = 100,250,500, 1000$, and $x_0 = x_{ff} \simeq (3\dod x_a^2)^{1/3}$. Lines opacity decrease with $\dod$ increase.}
	\label{3-motion-fig}
\end{figure}

\paragraph{Friction} $\gamma \sim {\rm Const}$ approximation can be used since particle has a large velocity during relatively small period compared to $t_{cap}$ as can be inferred from Figure~\ref{3-motion-fig}. Therefore we also take into account only gravity,
\be
w' - \gamma_0 t_0 w^2 - \frac{1}{4\rho^2} w^3 = 0,
\ee
for $w(\rho) \equiv t'(\rho) / t_0$ since $t'(\rho) = 1/\dot \rho(t)$, $t''(\rho) = - \ddot \rho / \dot \rho^3$. We introduce parameter $\epsilon \equiv \gamma_0 r_g \sim \dod x_a^2/x_0^3 \ll 1$ that is certainly small when particles cross event horizon independently (see Section~\ref{sec-accr}). Though, $\gamma_0 t_0 = \epsilon / 2\vk = \dod / 2x_0 \gtrsim 1$ at temperatures of interest. We expand solution, $w(\rho) = w^{(0)} + \epsilon w^{(1)}$, 
\be
{w^{(1)}}' - \frac{\epsilon}{\vk} w^{(0)} w^{(1)} - \frac{3}{4\rho^2}{w^{(0)}}^2 w^{(1)} = \frac{1}{2\vk} {w^{(0)}}^2,
\label{3-fric-eq}
\ee
with $w^{(0)} = -\sqrt{2} / \sqrt{1/\rho - 1/\rho_0}$. Since $\dot \rho(t_0) = 0$ we impose $w^{(1)}(\rho_0) = 0$ which seems to partially cancel divergence caused by ${w^{(0)}}^2$ presence in homogeneous equation solution,
\be
\lim_{\varepsilon\to0}\int_0^{w^{(1)}}\frac{du}{u+ \varepsilon} - \frac{3}{2}\int_1^{\rho/\rho_0}\frac{dq}{q(1-q +\varepsilon)} = \ln \frac{w^{(1)}}{f_1^3(\rho/\rho_0)} - \ln\varepsilon^{-1/2}.
\ee
Fortunately, remaining factor can be factored out leading to solution of homogeneous equation, $\exp{F} = f_1^3\left({\rho}/{\rho_0}\right) \exp{2\gamma_0 t^{(0)}(\rho)}$, and
\be
t^{(1)} = \int_{\rho_0}^{\rho} \left( e^{F(\rho')} + \frac{1}{2\vk} \int_{\rho_0}^{\rho'} e^{F(\tilde \rho)} {w^{(0)}}^2 d\tilde \rho\right) d\rho'.
\label{3-fric-sol}
\ee
Since $\vk \ll 1$ it is dominated by the second term. Exponent is $ \sqrt{2} \rho_0^{3/2} (f_1 + f_2) {\dod}/{x_0} \lesssim \frac{2}{3}\dist_0^{3/2} (f_1 + f_2)$ since equation of motion is applicable from $x_{ff} = (3 \dod x_a^2)^{1/3}$ (as is inferred from Figure~\ref{4-acc-regimes}). Therefore for relatively distant particles Laplace's method can be used to estimate the integral,
\be
t^{(1)} \sim \begin{cases}
	\displaystyle\frac{x_0^2}{\dod^2} f_1^3\left(\frac{\rho}{\rho_0}\right) e^{2 \gamma_0 t^{(0)}(\rho)},~~~\rho > \rho_p, \\
	\displaystyle\sqrt{\frac{\pi x_0}{\dod }} \frac{\rho_p^{3/2}}{\rho_0^{1/4}} \left( \rho_p + \sqrt{\frac{x_0^3 \rho_0^{1/2}f_{1p}}{\pi\dod^3\rho^3_p}}  - \rho\right) f_{1p}^{5/2}e^{2 \gamma_0 t^{(0)}(\rho_p)},~~~\rho < \rho_p
\end{cases} 
\ee 
with peak position $ \rho_p \simeq 2.5 \rho_0/(3+\dod\rho_0^{3/2}/\sqrt{2} x_0)$ and $f_{1p} \equiv f_1(\rho_p/\rho_0)$.

\section{\label{sec-accr}Accretion}

Any overdensity region accretes non-relativistic particles though we consider primordial black holes in homogeneous and isotropic Universe leading to motion with initial spherical symmetry. Newtonian gravity can be incorporated to describe it if the asymptotic mass of accreting particles is much lesser than of black hole~\cite{Malec-1999}, $\int_{r_g}^{r_H} m n r^2 dr \ll M$. With freeze-out yield $Y_{fo}$ it translates into
\be
x \lesssim \left( \frac{3}{4 Y_{fo}} x_a^2\right)^{1/3} + \frac{x_a^{4/3}}{x}
\ee
for $Y_{fo} \sim Y_{rel}$ we obtain condition on appearance time $x_a \gtrsim (x_{in}/6.6)^{3/2} \sim 0.3$. Afterwards particles develop an overdensity profile and this condition is softened therefore we do not include particles effects in study. Validity of non-relativistic approach to heavy particles accretion was also mentioned in ref.~\cite{Lora-Clavijo-2014}.

\begin{figure}
	\centering
	\includegraphics[width=0.49\textwidth]{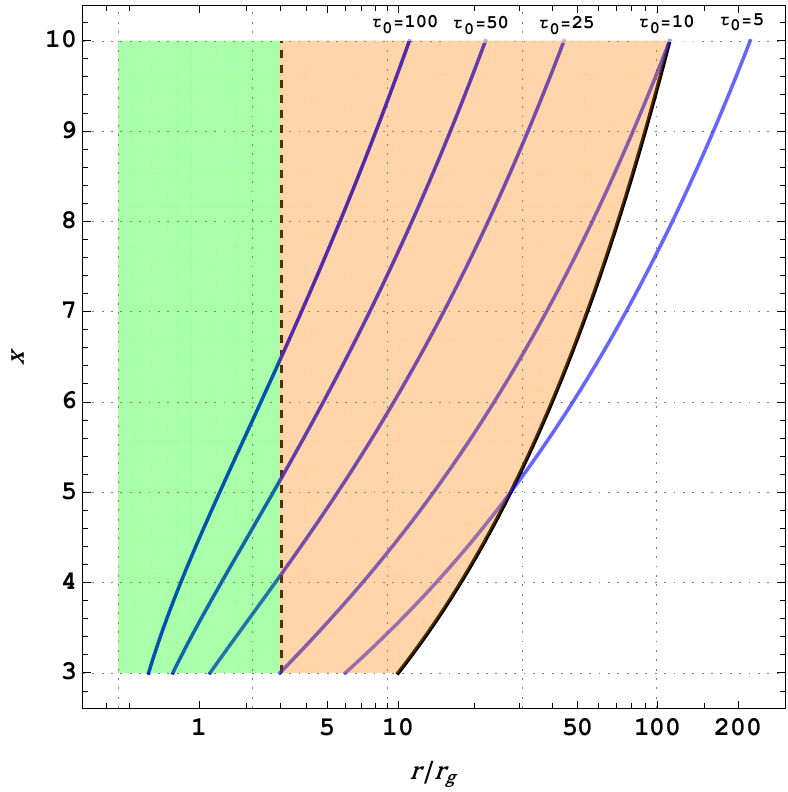}
	\includegraphics[width=0.49\textwidth]{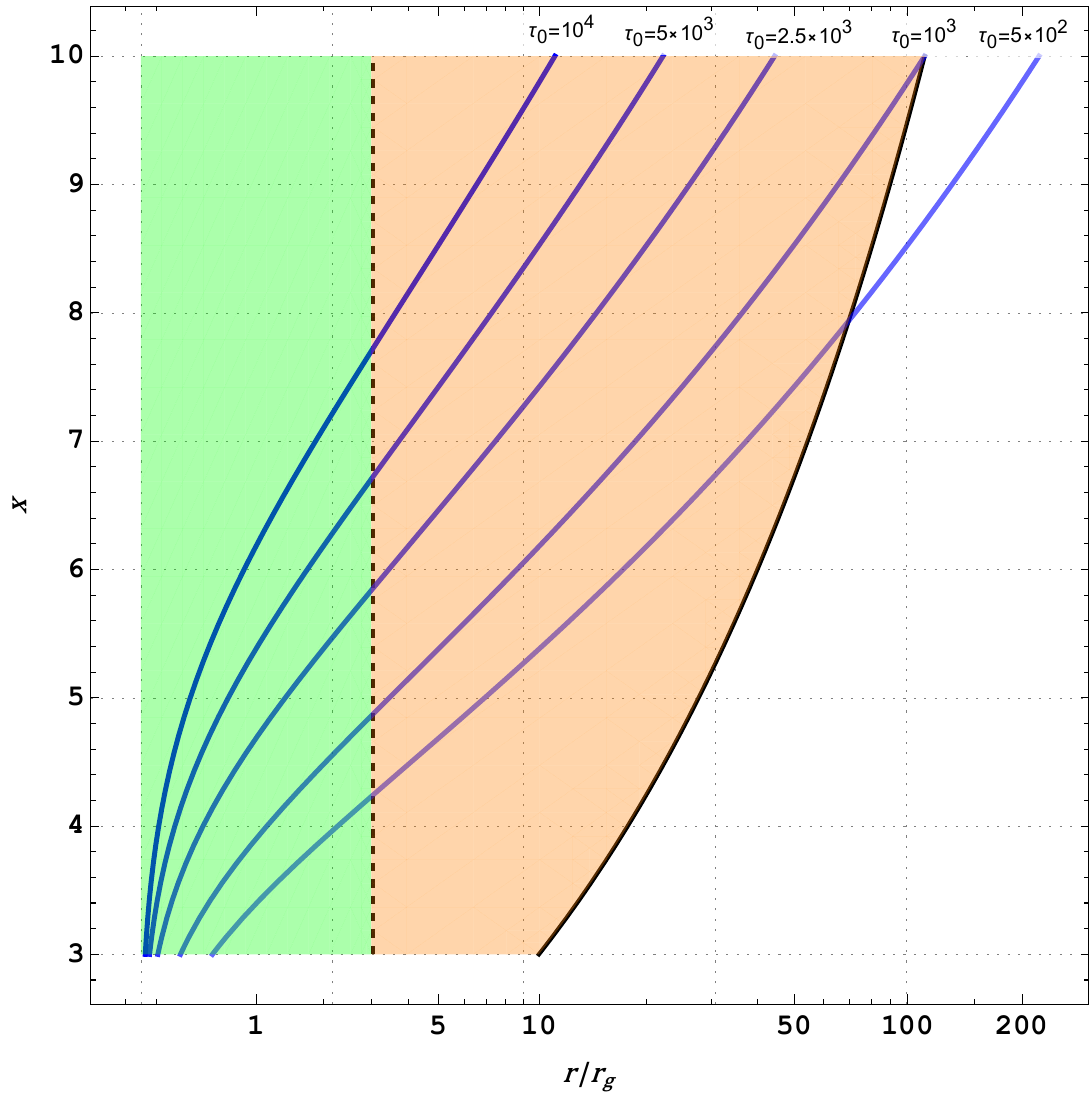}
	\caption{Evolution of particles horizon (black line) and mean free paths (blue lines) depending on $\dod$. Regions inside are separated by accretion radius (dashed line). Left: case of geometrical cross-section $\braket{\sigma v} = \cpl^4/m^2$. Right: case of fermion-fermion scattering cross-section at low energies $\braket{\sigma v} = \cpl^4/m^2x^2$. Pictures are provided for $x_a = 0.95$.}
	\label{4-acc-regimes}
\end{figure}

Characteristic accretion length,
\be
r_{acc} \equiv \frac{2GM}{c_{s,\infty}^2} = 6(1+R)GM,
\ee
with sound speed of relativistic particles, $c_s = 1/\sqrt{3}$, and $R = (\rho_{1 + \bar 1} + \press_{1+ \bar 1})/(\rho_{rel} + \press_{rel}) = 3\rho_{1+\bar 1}/4\rho_{rel}$ where $\press$ is pressure. It is bounded by $\dist_{acc} \simeq 3 \dist_g  \left( 1 + R(\dod)\right) \lesssim (3 + 0.04x) \dist_g$ and in last equality we put relativistic number density of $N_1$ with $g_1 = 1$. Whether particles fall inside primordial black hole {\it independently} or {\it coupled} to relativistic plasma depends on relation of mean free path to accretion radius~\cite{FrKiRa-Accr-Power}. This comparison is provided in Figure~\ref{4-acc-regimes} from which we conclude that pressure affects motion only initially, until $x_{ff} = (3\dist_g \dod x_a^2)^{1/3}$. Afterwards non-relativistic particles interaction with plasma reduces to friction. When $\dist_{free} = \dist_H$ or $x_f = \dod$ non-relativistic particles begin to freely propagate throughout the Universe. Note, that $\dod = {\rm Const}$ on left picture while $\dod \propto 1/x^2$ on the right as expected in the case of fermion on fermion scattering via Yukawa interaction. The latter requires larger $\dod$ for the same $x_f$. Moreover, pressure influence on motion takes more time. Both factors reduce accretion rates making fermion on fermion scattering less appealing.

Accretion begins from two-component liquid if $\dod > 9 / x_a^2$ ($\dod > 81 / x_a^2$ for $\dod \propto x^{-2}$). This regime is characterized by Bondi radius, $r_B = GM/2c_{s,\infty}^2$, at which particle speed exceed the local sound speed. Quasi-stationary Bondi limit is applicable if cosmological parameters change is negligible on characteristic timescale, $t_{cr} \sim r_B / c_{s,\infty} < t_H$~\cite{Ricotti-2007}. However, instead of highly non-relativistic electrons accretion we consider accretion proceeding with scattering and annihilation processes. At relevant times $x < x_{ff}$ the latter can significantly affect the vicinity of black hole diminishing role of initial stage of accretion. This argument is further strengthened by relativistic fluid pressure behavior near the event horizon $\press = \rho(t) \left((1 - r_g/r)^{-1/2} - 1\right) $~\cite{Nandra-2011}, though they obtain this expression imposing zero pressure at infinity.

\paragraph{Independent particles accretion}

We adopt Lagrangian approach where fluid particle is marked by its initial position $r_0$ and trajectory is $r = r(t, r_0)$. If particles are conserved in differential volume then $nr^2dr = n_\infty r_0^2 dr_0$, or
\be
n(r_g) = \left. n_\infty \frac{r_0^2}{r_g^2} \middle/  \frac{\pd r}{\pd r_0}\right|_{t=t_{cap}} = n_\infty \phi(t_{cap}, r_0).
\label{4-ndens-profile}
\ee
Equation $r(r_0, t_{cap}) = r_g$ fixes a unique correspondence between $t=t_{cap}$ and $r_0$. Therefore we consider $r_0 = r_0(t)$, dependence of initial position on capture time. Then identity $v(r_g) \equiv \left.{\pd r} / {\pd t} \right|_{t=t_{cap}}$ allows to transform particles absorption rate by black hole is $\dot N_{abs}(t) = 4\pi r_g^2 n(r_g, t) \, v(r_g, t)$ into
\be
\dot N_{abs} = \left. 4\pi r_0^2 n_\infty \middle/ \left.\frac{\pd t(r, r_0)}{\pd r_0}\right|_{r=r_g} \right. .
\label{4-Nabs-expr}
\ee
We apply this formula for protons in molecular cloud affected only by gravity assuming steady flow, i.e. $r_g / r_0 \ll 1$. Then $ct_{cap} \simeq {\pi}/{2} \cdot \sqrt{r_0^3/r_g} = ct_0$ and $\dot N_{abs} = {32}/{3\pi} \cdot r_g c^2 t n_\infty$. Mass evolution,
\be
M = M_0 \times {\rm exp} \left\lbrace 4 \cdot 10^{-15} \frac{m}{\rm GeV} \frac{n_\infty}{10^8 \, {\rm m^{-3}}} \left( \frac{t}{\rm yr} \right)^2 \right\rbrace,
\ee
requires few billion years of accretion to appreciably increase mass being in agreement with~\cite{ZelNov1971}. Free fall approximation of protons accretion is valid for small black holes~\cite{DolRud-2023} while at some point of mass growth pressure effects should come into play. 

In expanding Universe we consider $Y(x) = n/s$ therefore equation~(\ref{4-Nabs-expr}) becomes
\be
\frac{\pd \dist_g(x)}{\pd x} = 4 \dist_0^2 Y_\infty \left.\frac{x_a^4}{x^2} \middle/ \frac{\pd x^2}{\pd \dist_0}\right|_{\dist=\dist_g}
= 4 \dist_0^2 \left.\frac{Y_\infty}{x_a^2\cdls^3} \middle/ \frac{\pd x}{\pd \dist_0}\right|_{\dist=\dist_g},
\ee
where $\cdls = x/x_a^2$ and capture time
\be
x_{cap} \approx x_0 \sqrt{ 1 + \frac{x_a^2}{x_0^2} \pi \sqrt{\frac{\dist_0^3}{\dist_g}} \left[ 1-  0.36 \frac{\dist_0^3 x_a^4}{\dist_g x_0^4}\right]}.
\ee
Now $\dist_0$ approximately depends on time as
\be
\dist_0(x) \sim \left[\frac{\dist_g(x^2-x_0^2)^2}{\pi^2 x_a^4}\right]^{1/3} \left( 1 + \frac{0.24 (x^2-x_0^2)^2}{\pi^2 x_0^4 - 1.08 (x^2-x_0^2)^2}\right).
\ee
Following motion equation solution we set $x_0 = x_{in}$ for $\dist_0 \leqslant \dist_e(x_{in})$ while $x_0 \sim x/\sqrt{1+0.82\pi/\sqrt{2}}$ for larger $\dist_0$ or $x > x_{in}\sqrt{1+0.82\pi/\sqrt{2}}$.

\begin{figure}
	\centering\includegraphics[width=.45\textwidth]{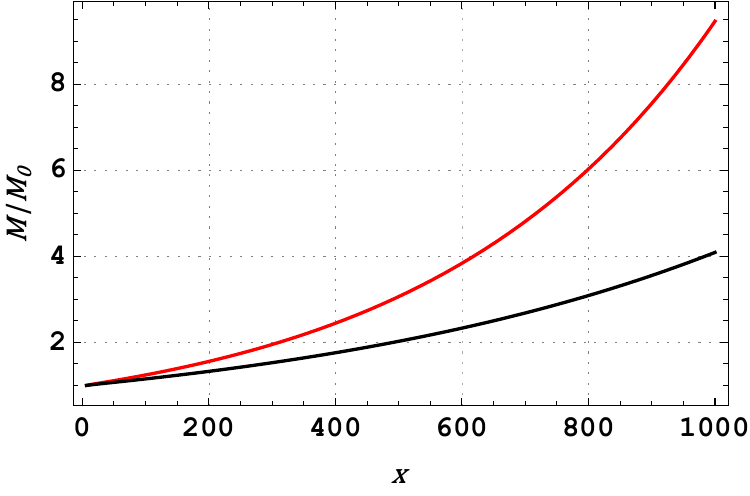}\includegraphics[width=.45\textwidth]{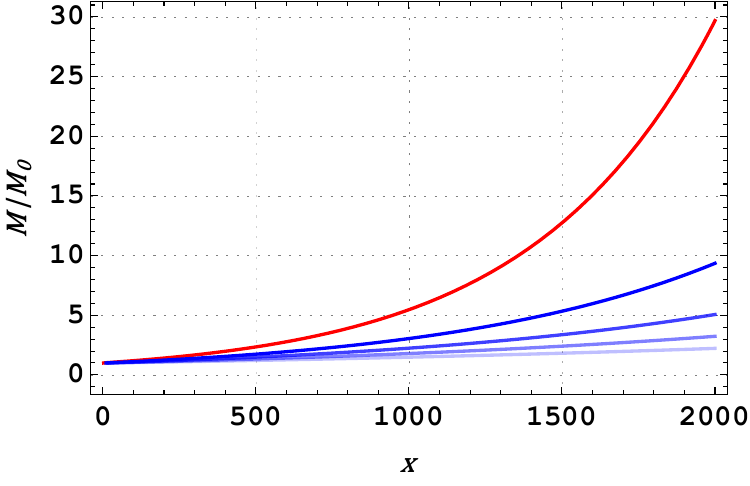}
	\caption{Left picture: $M/M_0$ evolution for black hole surrounded by non-relativistic gas remaining relativistic number density in the early Universe (red) and in present Universe (black). Right picture: $M/M_0$ evolution for the freeze-out case with $\dod = 100,250,500,1000$.}
	\label{4-accr-rates-figure}
\end{figure}

We evaluate accretion until $x_{cl} =2000$ which correspond to $\abnc \sim 10^{-9}$ and $x_a = 2$. Primordial black holes growth faster than in present Universe, as shown in left picture in Figure~\ref{4-accr-rates-figure}, and can grow significantly given sufficient time. It depends upon the freeze-out value $Y_{fo} = Y_{fo}(\dod)$ as shown in right picture in Figure~\ref{4-accr-rates-figure}. Second, friction affects accretion resulting in further growth delay. Though we do not discuss annihilation in the vicinity of black hole, it also increases such delay. 

\begin{figure}
	\centering\includegraphics[width=.45\textwidth]{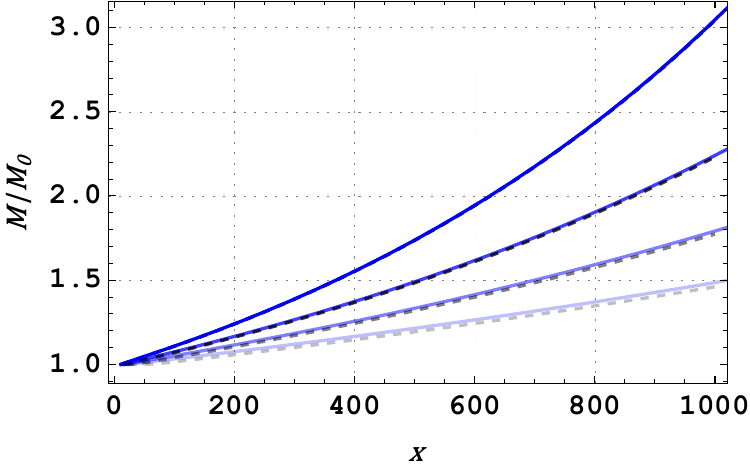}\includegraphics[width=.45\textwidth]{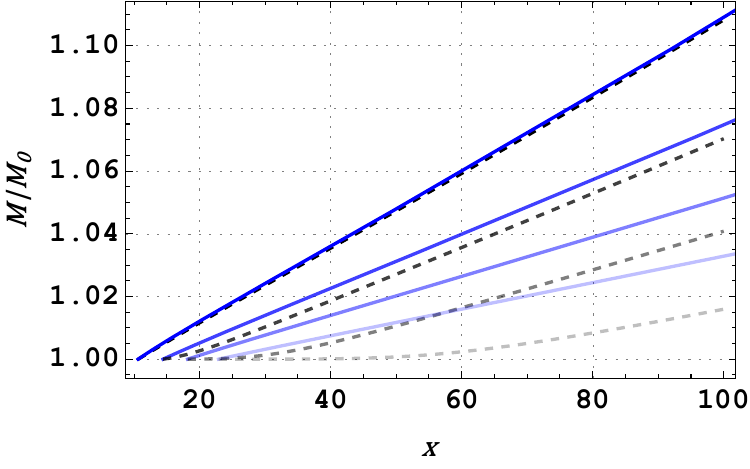}
	\caption{PBH growth rates for $\dod = 100,250,500,1000$. $\dod$ influence on freeze-out number density and $x_f$ is taken into account for blue lines. Black dashed lines also include friction term.}
	\label{4-rates-fric}
\end{figure}

\paragraph{Asymmetry generation}

\begin{figure}
	\centering\includegraphics[width=.45\textwidth]{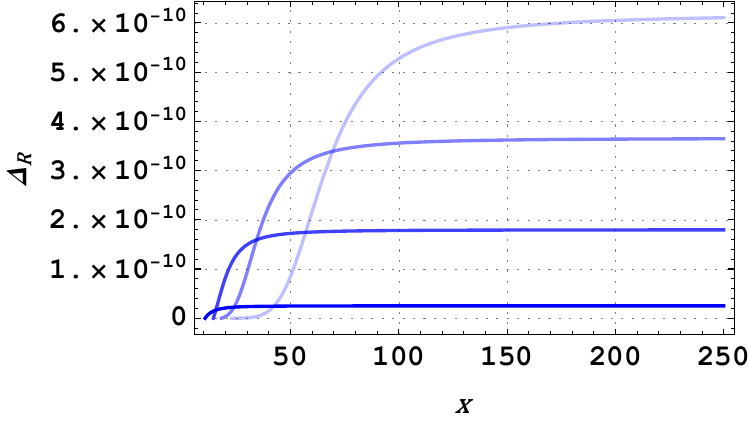}\includegraphics[width=.45\textwidth]{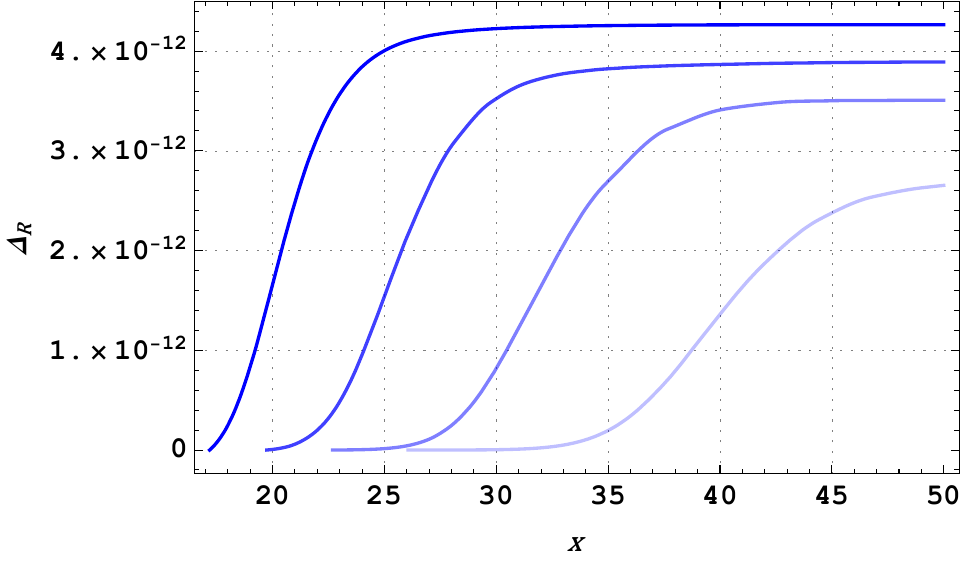}
	\caption{Asymmetry evolution for $\abnc = 10^{-5}$. Left: case of geometrical cross-section with $\dod = 100,250,500,1000$. Right: case of fermion on fermion Yukawa scattering with $\dod \times 10^{-5} = 1.25, 2.5, 5, 10$.}
	\label{4-ASMR-evol}
\end{figure}

In ref.~\cite{Dolgov-2020} an order of magnitude estimation for CPV magnitude was provided, $\varepsilon_s' \sim \cpl^2$ based on  analogy with baryogenesis via decay processes. Calculation for fermion on fermion $N_1 + \bar a \to N_1 + \bar c$ Yukawa interactions revealed additional time dependence $1/x^{3/2}$~\cite{Ambrosone-2022},
\be
\varepsilon_s' \equiv \frac{\sigma(N_1 + \bar a \to N_1 + \bar c) -\sigma(\bar N_1 + a \to \bar N_1 + c)}{\sigma(N_1 + \bar a \to N_1 + \bar c) + \sigma(\bar N_1 + a \to \bar N_1 + c)} \lesssim \frac{\cpl^2}{\sqrt{2}\pi x^{3/2}}.
\label{4-CPV-par}
\ee
maximized if $|m_b - m_Y| \ll T$, $m_X \simeq m_\psi$ and does not depend upon $m_\phi$ choice so we put $m_\phi = 0$. However, corresponding cross-section $\sigma_{ff} \sim \cpl^4 (1-m^2/s)^2/s$ by expanding $s \sim m^2 + 2mT$ leads to thermally averaged cross secton $\braket{\sigma_s v} \sim \cpl^4/m^2 x^2$. As expected fermion on fermion $2\to2$ process has a threshold and is suppressed at low temperatures (as discussed in for instance~\cite{Birdman-Lect}). It also means that annihilation processes, such as $N_1 + \bar N_1 \to a \bar c$ possessing additional phase factor $ |\vec p'| / |\vec p| \sim \sqrt{1-4m^2/s}$ are also suppressed at low temperatures.

Therefore friction difference obtains not $1/x^{3/2}$ but $1/x^{7/2}$ factor therefore fermion on scalar $N_1 + \phi \to N_1 + \psi$ scattering may become more favorable especially if scattering bubble-diagram can be resonantly enhanced by analogy with resonant leptogenesis~\cite{Pilaftsis-2003}, while CPV magnitude connection between neutrino decay and scattering processes is discussed in~\cite{Buchmuller-1997}.

Following~(\ref{4-CPV-par}) we represent $\gamma = \gamma_{av}\pm \delta\gamma$ with different signs corresponding to particles and antiparticles. It enters solution~(\ref{3-fric-sol}) via $\epsilon \equiv \gamma r_g \sim \dod x_a^2 / x^3$. Then accretion rates difference is
\be
\frac{\pd N_R}{\pd x} = \frac{2\varepsilon' \epsilon}{1 + \epsilon \xi^{(1)} / \xi^{(0)}} \frac{\xi^{(1)}}{\xi^{(0)}} \frac{\pd N}{\pd x},
\ee
with $\xi^{(i)} \equiv \pd {x^{(i)}}^2/\pd \dist_0$ and $N_R$ being some quantum number assigned to $N_1$ particles, for example, $B-L$ since we discuss leptogenesis.

Integration over PBH population divided by entropy density $s$ leads to expression for asymmetry $\Delta_R$ evolution. In case of monochromatic spectrum it takes a simple form,
\be
\frac{\pd \Delta_R}{\pd x} = \frac{2.7\abnc}{x_a} \frac{2\varepsilon' \epsilon}{1 + \epsilon \xi^{(1)} / \xi^{(0)}} \frac{\xi^{(1)}}{\xi^{(0)}} \frac{\pd \dist_g}{\pd x}.
\ee
Asymmetry evolution for chosen parameters is provided in figure~\ref{4-ASMR-evol}. It was shown for geometrical cross-section that using small $\dod$ parameter resulting asymmetry increases with $\dod$ increase. However, for fermion on fermion cross section $\dod$ has generally higher value and for chosen parameters we find asymmetry decrease with $\dod$ increase. Since $x_{ff} \propto \dod^{1/3}$, larger $\dod$ also increases time for asymmetry generation but not significantly.

\section{\label{sec-disc}Discussion}

Though today Universe expansion scarcely affects local dynamics in the early times situation is quite the opposite. Taking it into account at the lowest order, $-qH^2r$, we found capture time in near region, $r_0 < r_e(t)$ to decrease compared to gravity while distant particles $r_0 \gtrsim 2 r_e(t)$ follow the Hubble flow. In the intermediate region we expect motion to change from recession to accretion assuming that they start motion towards black hole since $r_e(t) = r_0$. It leads to delay in distant particles capture reducing effect of expansion. However, analytical solution would allow to make a thorough conclusion.

Nevertheless, growth rates in the early Universe are faster than in gravity. We note, that taking into account effect of matter overdensity near black hole would have a larger effect on the early Universe accretion since it also increases equilibrium radius $r_e$. 

We argue that despite accretion begins from two-component liquid it has negligible effect on both black hole mass and produced asymmetry. Therefore we consider evolution from $x_{ff} = (3 \dod x_a^2)^{1/3}$ until $x_{cl} = \abnc x_a$ starting from independent capture of particles. After $x_{cl}$ we assume that spherical symmetry is likely to be broken and clustering should be taken into account making numerical studies possibly the only approach.

Fortunately, asymmetry production proceed in a narrow range of time after $x_{ff}$ allowing to consider analytical solution. It secludes prolonged scenarios of baryogenesis which can be expected because friction effect in the early Universe weakens with time. Since we consider $x > x_{ff}$ we can treat friction perturbatively.

Herein we only discuss asymmetry production associated with $N_1$-particles quantum number. It can be either transferred into the SM sector via $N_1$'s decays or by virtue of annihilation processes in the vicinity of black hole because at infinity particles are assumed to depart from chemical equilibrium. Since for the simplest monochromatic mass-spectrum of PBHs $\Delta_R \propto \abnc$ it is preferable to consider larger $\abnc$. Therefore if not all then most of PBHs should evaporate in order to escape constraints on cold DM parameter density. Therefore it is interesting to study whether PBHs can evaporate in clusters. However, PBHs evaporation can dilute produced asymmetry~\cite{Chaudhuri}.

The further development of leptogenesis mechanism is incorporation of annihilation processes and asymmetry transfer study. It would also soften parameter space constraints that was not included in present study.

\acknowledgments

The work is supported in the framework of the State project
``Science'' by the Ministry of Science and Higher Education of the
Russian Federation under the contract 075-15-2024-541.

\end{document}